\documentclass[fleqn,11pt]{article}

\usepackage{amsmath,latexsym,amssymb,cite}
\usepackage{graphicx}
\usepackage{hyperref}

\usepackage{amsfonts,amssymb,cite}
\usepackage{graphicx}

\def\noi{\noindent}

\newcommand{\Title}[1]{\noi {{\Large\bf #1}}\\[1ex]}

\newcommand{\Author}[2]{\noi{\bf #1}\\[2ex]\noi{\normalsize\it #2}\\}

\newcommand{\Abstract}[1]{\vskip 2mm \begin{center}
        \parbox{16.4cm}{\small\noi #1} \end{center}\medskip}

\newcommand{\foom}[1]{\protect\footnotemark[#1]}

\def\nqq{\hspace*{-2em}}
\def\nhq{\hspace*{-0.5em}}

\def\cm{\hspace*{1cm}}

\def\ten#1{\mbox{$\times 10^{#1}$}}

\def\Jl#1#2{#1 {\bf #2},\ }

\def\ApJ#1 {\Jl{Astroph. J.}{#1}}
\def\CQG#1 {\Jl{Class. Quantum Grav.}{#1}}
\def\DAN#1 {\Jl{Dokl. AN SSSR}{#1}}
\def\GC#1 {\Jl{Grav. Cosmol.}{#1}}
\def\GRG#1 {\Jl{Gen. Rel. Grav.}{#1}}
\def\JETF#1 {\Jl{Zh. Eksp. Teor. Fiz.}{#1}}
\def\JETP#1 {\Jl{Sov. Phys. JETP}{#1}}
\def\JHEP#1 {\Jl{JHEP}{#1}}
\def\JMP#1 {\Jl{J. Math. Phys.}{#1}}
\def\NPB#1 {\Jl{Nucl. Phys. B}{#1}}
\def\NP#1 {\Jl{Nucl. Phys.}{#1}}
\def\PLA#1 {\Jl{Phys. Lett. A}{#1}}
\def\PLB#1 {\Jl{Phys. Lett. B}{#1}}
\def\PRD#1 {\Jl{Phys. Rev. D}{#1}}
\def\PRL#1 {\Jl{Phys. Rev. Lett.}{#1}}

\def\al{&\nhq}
\def\lal{&&\nqq {}}
\def\eq{Eq.\,}

\def\beq{\begin{equation}}
\def\eeq{\end{equation}}
\def\bear{\begin{eqnarray}}
\def\bearr{\begin{eqnarray} \lal}
\def\ear{\end{eqnarray}}
\def\earn{\nonumber \end{eqnarray}}
\def\nn{\nonumber\\ {}}

\def\nnn{\nonumber\\ \lal }
\def\nnnv{\nonumber\\[5pt] \lal }

\def\yyy{\\[5pt] \lal }
\def\eql{\al =\al}

\def\dst{\displaystyle}

\def\fracd#1#2{{\dst\frac{#1}{#2}}}

\def\Half{{\fracd{1}{2}}}

\def\e{{\,\rm e}}
\def\d{\partial}

\def\sign{\mathop{\rm sign}\nolimits}
\def\diag{\mathop{\rm diag}\nolimits}

\def\const{{\rm const}}

\def\then{\ \Rightarrow\ }

\def\mn{_{\mu\nu}}
\def\MN{^{\mu\nu}}
\def\mN{_\mu^\nu}

\def\ssph{static, spherically symmetric}

\def\Scz{Schwarz\-schild}

\def\GN{G_{\mathrm{N}}}
\def\PhiN{\Phi_{\mathrm{N}}}

\begin{document}
\twocolumn[

\Title{On variations of $G$ in the geometric scalar theory of gravity}

\Author{K. A. Bronnikov\foom 1}
	{\small
	Center fo Gravitation and Fundamental Metrology, VNIIMS, 
		Ozyornaya ul. 46, Moscow 119361, Russia;\\
	 Institute of Gravitation and Cosmology, RUDN University, 
		ul. Miklukho-Maklaya 6, Moscow 117198, Russia;\\
	National Research Nuclear University ``MEPhI'', 
		Kashirskoe sh. 31, Moscow 115409, Russia}

\Abstract
	{We analyze the possible variability of the effective Newtonian gravitational constant 
	$\GN$ in space and time in the framework of the geometric scalar theory of gravity 
	suggested by M. Novello et al. [JCAP 06, 014 (2013); arXiv: 1212.0770]. 
	Spatial variations of $\GN$ in the Solar system are shown to have orders of magnitude 
	detectable by modern instruments. As to variations of $\GN$ with  cosmological time.  
	it is shown (at least for the particular formulation of the theory discussed in the 
	original paper and the corresponding cosmological models) that these variations
	are more rapid than is allowed by observations.       
       }

] 

\section{Introduction}

  General relativity (GR) is well known to have a brilliant experimental status with respect 
  to local phenomena, confirmed by observations in the Solar system and on other 
  astrophysical objects on stellar or sub-galactic scales. The most important experimental 
  achievements of the recent years, the detection of gravitational waves and observations
  with the Event Horizon Telescope, apparently confirms GR predictions for extremely 
  strong gravitational fields, the existence and properties of black holes, see, e.g., 
  the recent reviews \cite{GW, EHT}.  
  
  Nevertheless, there is almost a universal opinion of the theorists that GR is
  not an ultimate theory of gravity and needs an improvement, and an enormous 
  number of its extensions and modifications are discussed and analyzed 
  in the physical literature, see, e.g., \cite{modif1, modif2} and references therein.
  The reasons for such views are twofold. On one hand, beginning with the 
  galactic length scale, and especially in cosmology, the observational status 
  of GR is not so perfect due to the well-known Dark Matter (DM) and Dark Energy
  (DE) problems, the first one above all related to missing mass in galaxies and
  clusters of galaxies, the second one to the observed accelerated expansion of the 
  Universe \cite{DM1, DM2, DE1, DE2, DE3}. Another group of reasons is connected 
  with the problems inherent to the theory itself. Thus, the most important solutions
  of GR contain space-time singularities with diverging values of curvature 
  invariants,  indicating situations where the theory cannot work any more. 
  There are also long-standing problems with quantization of gravity and with 
  its unification with other physical interactions. Such a wide set of problems 
  has caused the advent of a great diversity of extended, or alternative theories
  of gravity.
  
  Some of them differ from GR by inclusion of additional fields (scalar-tensor, 
  Einstein-aether, bimetric, tensor-vector-scalar (TeVeS) theories, etc.; others, such
  as, for instance, $f(R)$ and many more complex theories contain higher-order 
  derivatives of the metric tensor. Numerous theories involve small or large extra 
  dimensions (Kaluza-Klein type or brane-world theories, respectively), some of 
  the latter being related to the string concept, and many theories make use of 
  non-Riemannian geometries, e.g., Finsler models, models with torsion 
  and/or nonmetricity --- see the vast bibliography in \cite{modif1,modif2}.
  
  A common feature of all such models is that they introduce new dynamic 
  degrees of freedom as compared to GR. The Geometric Scalar Theory of 
  Gravity (GSG), recently proposed by Mario Novello and co-authors 
  \cite{no12}, makes a step in the opposite direction and shows that some
  opportunities of interest are not yet exhausted in attempts to simplify the 
  description of gravity instead of adding its complexity. As said in \cite{no16},
  ``Here we propose that it may be interesting to take a huge step backward 
   and explore models in which the gravitational degrees of freedom are
   just described by the field $\Phi$.''
   
  GSG is a metric theory of gravity in which all kinds of nongravitational
  matter interact with the dynamic field $\Phi$ only through the gravitational
  metric $q\mn$. In addition to the gravitational metric $q\mn$, GSG also
  employs the auxiliary Minkowski metric $\eta_{\mu\nu}$ which is not observable.
  It turns out that this kind of theory has a chance to be viable, unlike previous
  attempts to build a relativistic scalar theory of gravity (see
  \cite{nord, finn} and detailed discussions in \cite{no12, no16}). 
  Thus, by properly choosing the parameters of the theory in such a way 
  that $q\mn$ for a field of a gravitating center has the \Scz\ form, it becomes 
  possible to reproduce Newton's theory in the weak field limit and 
  all local classical effects of GR \cite{no12}. In cosmology, GSG has been 
  shown \cite{no14} to be able to solve the singularity, horizon and flatness 
  problems without appeal to exotic kinds of matter, in particular,
  predicting a bounce instead of a singularity; it is also shown to present 
  a basis for structure formation by gravitational instability \cite{no14}. 
  A study of gravitational waves in GSG \cite{toni19} has shown that 
  they are described in the weak field approximation of this theory in a way
  similar to GR, they propagate at the same speed as light, but a 
  characteristic longitudinal polarization mode, absent in GR, is predicted,
  so its possible discovery can be a strong evidence in favor of GSG. 
  All these features seem to make GSG one of the promising alternatives 
  to GR.
  
  The present study shows, however, that GTG faces a serious problem 
  with too large variations of the effective Newtonian gravitational constant $\GN$.
  Such variations are predicted by many non-Einsteinian theories of gravity,
  see, e.g., \cite{G1, G2, G3, G4, G5, G6, G7} and references therein. Meanwhile,
  there are strong observational constraints on time variations of $\GN$ 
  \cite{Gt1, Gt2, Gt3} and milder ones on its spatial dependence, mostly 
  related to the so-called fifth force concept \cite{G1}.  
  Our purpose here will be to estimate both temporal and spatial variations of $\GN$
  in the version of GSG presented and discussed in \cite{no12, no14, no16, toni19}.  
  
  The paper is organized as follows. The next section outlines some basic 
  relations of GSG. Section 3 is devoted to finding a general expression for the
  local value of $\GN$ in terms of the fundamental scalar field $\Phi$, and in
  in Section 4 it is used for obtaining the relevant estimates. Section 5 contains
  some concluding remarks.  

 \section {Basic relations of GSG}
  
  As mentioned above, GSG contains two metrics, the observable one, $q\mn$,
  and the flat auxiliary one, $\eta\mn$, and both can be used in an arbitrary 
  coordinate system since the theory is generally covariant, free of any privileged 
  reference frame. The two metrics are connected by a disformal transformation
  described by the relations\footnote
  	{We use the metric signature $(+\ -\ -\ -)$; the curvature tensor is
  	defined via Chrisfoffel symbols for $q\mn$ as
        $R^{\sigma}{}_{\mu\rho\nu} =\d_\nu\Gamma^{\sigma}_{\mu\rho}-\ldots$,\ 
  	$R\mn = R^{\sigma}{}_{\mu\sigma\nu}$, so that the scalar $R = R_\mu^\mu > 0$
  	for de Sitter space-time or the matter-dominated cosmological epoch; the
        system of units $c = \hbar = 1$.}  
\bearr                               		\label{q^}
	q^{\mu\nu} = \alpha \, \eta^{\mu\nu} + \frac{\beta}{w} \,
	\d^{\mu}\Phi \,\d^{\nu} \Phi,
\yyy                                            \label{q_}
	q_{\mu\nu} = \frac{1}{\alpha} \, \eta_{\mu\nu} -
	\frac{\beta}{\alpha \, (\alpha + \beta) \, w} \, \d_{\mu} \Phi
	\, \d_{\nu} \Phi.
\ear
  where $ \d^{\mu} \Phi \equiv \eta^{\mu\nu}\,\d_{\nu} \Phi$, the parameters
  $\alpha >0$ and $\beta$ are certain dimensionless functions of $\Phi$, and
\bearr
	w \equiv \eta\MN \d_{\mu}\Phi \d_{\nu} \Phi,                  \label{w}
\nnnv
	\Omega \equiv q\MN \d_{\mu}\Phi \d_{\nu} \Phi = (\alpha + \beta) \,w.
\ear

  The dynamics of the theory is specified by the action $S = S_g + S_m$,
  where $S_g$ and $S_m$ are its gravitational and matter parts,
  respectively:
\bearr                                                   \label{S_g}
	S_g = \frac{1}{16\pi G_0}\int \sqrt{-\eta} \,d^4 x\, V(\Phi)\, w,
\yyy                                                     \label{S_m}
	S_m = \int \sqrt{-q}\, d^4 x\, L_m.
\ear
  Here, $\eta = \det (\eta\mn)$, $q = \det (q\mn)$, $V(\Phi) > 0$ is a certain
  function called the potential, $L_m$ is the Lagrangian of matter, and
  $G_0$ is an initial constant introduced similarly to the usual
  gravitational constant in GR, but, in general, it does not coincide with
  Newton's constant of gravity in GSG, as we shall see below. Furthermore,
  it is important that the stress-energy tensor of matter $T\mn$, defined in
  the usual way in terms of $q\mn$,
\beq                                                            \label{SET}
	T_{\mu\nu} \equiv  \frac{2}{\sqrt{- q}} \,
	\frac{\delta( \sqrt{- q} \, L_{m})}{\delta q^{\mu\nu}},
\eeq
  obeys the conservation law $\nabla_\nu T\mN =0$ (again in terms of $q\mn$).

  The functions $\alpha(\Phi)$, $\beta(\Phi)$ and $V(\Phi)$, entirely fixing
  the formulation of the theory, are specified in \cite{no12} in such a way
  that the vacuum field equation for $\Phi$ has the form $\Box \Phi =0$,
  where $\Box = q\MN \nabla_\mu\nabla\nu$ is the d'Alembert operator
  corresponding to the metric $q\mn$, and the \ssph\ vacuum solution for
  $\Phi$ leads to the Schwarzschild form of the metric $q\mn$.
  It is this circumstance that makes the weak-field predictions of GTG
  (e.g., in the Solar system) the same as those of GR. Specifically,
  we have \cite{no12}
\bearr
	\alpha + \beta = \alpha^3 V(\Phi),        	\label{func}
\nnnv
	4\alpha^3 V = (\alpha -3)^2,\qquad  \alpha = \e^{-2\Phi}.
\ear

  The field equation for $\Phi$ can be written in the following way for the
  case where $L_m$ describes a perfect fluid with density $\rho$ and
  pressure $p$:
\beq                                                 \label{eq1}
	\sqrt{V} \,\Box\, \Phi = - \frac{\kappa}{2}
	\left(\frac{2\alpha}{\alpha-3} \, \rho - 3 p \right).
\eeq
  In terms of the Minkowski metric $\eta\mn$, using (\ref{q^}), we obtain
\beq                                                   \label{eq2}
	\Box_M\Phi + \Half \frac{V_\Phi}{V} w = \frac{\kappa}{2}\,
	\left(\frac{1}{\alpha \sqrt{V}}\right)^3
	\left(3p-\frac{2\alpha}{\alpha-3}\rho\right),
\eeq
  where $\kappa = 8\pi G_0$ and $V_\Phi \equiv dV/d\Phi$. (This equation
  coincides with \eq (46) of \cite{no12}.) Or equivalently, due to (\ref{func}),
\beq                                                   \label{eq3}
	\Box_M\Phi + \frac{\alpha-9}{\alpha-3} w =
	\frac{\kappa}{\alpha^3 V^{3/2}}
	\biggl(\frac{3}{2}p - \frac{\alpha}{\alpha-3} \rho\biggr).
\eeq

\section{The effective gravitational constant}

  Let us now find out the expression for the effective Newtonian
  gravitational constant $\GN$ in the weak-field and low-velocity limit of
  GSG. In this limit, the Newtonian gravitational potential $\PhiN$
  due to a matter distribution with density $\rho$ should, as usual,
  satisfy the Poisson equation
\beq                                               \label{Poisson}
	\Delta \PhiN = 4\pi \GN \rho,
\eeq
  where $\Delta$ is the flat-space Laplace operator.

  The Newtonian approximation should be valid for the gravitational
  interaction of any system of bodies located so closely to each other that
  the space-time curvature could be neglected, and moving with very small
  relative velocities, so that in a suitable reference frame and in properly
  chosen coordinates the gravitational field is dominated by the temporal
  component of the space-time metric written as
\beq
	g_{00} = 1 + 2\PhiN, \cm |\PhiN| \ll 1,       \label{Newt}
\eeq
  where $\PhiN$ obeys the equation (\ref{Poisson}). Such a small domain
  can be considered as a close neighborhood of a certain 4D point $x_0^\mu$
  in which the scalar field $\Phi$ changes so slowly that $\Phi(x_0^\mu)$
  can be taken as a constant background value, while $\PhiN$ should be
  related to comparatively rapidly changing deflections from this constant
  value. In other words, $\Phi$ should be taken in the form
\bearr
	\Phi (x^\mu) = \Phi_0 + \Phi_1 (x^\mu), \quad\    \label{Phi1}
\nnnv	
	\Phi_0 = \Phi (x_0^\mu) \approx \const.
\ear
  Our task is to find a relationship between $\Phi_1$ and $\PhiN$ and to
  bring \eq (\ref{eq3}) to the form (\ref{Poisson}) (where we put $p=0$
  since we consider nonrelativistic matter) by substituting there the
  expansion (\ref{Phi1}).

  Let us note that examples of physical situations where such an
  approximation is quite plausible include gravitational fields in various local
  systems (a galaxy, a stellar cluster or a planetary system) against a slowly 
  varying cosmological or larger-scale background which can be regarded 
  constant on the time or length scales small as compared to the corresponding 
  scales of the background. Such examples are (i) the dynamics of a galaxy 
  against the time scale of the Universe, (ii) the dynamics of a planetary system 
  against the background of the galactic gravitational field, which is almost 
  time-independent and very slowly varies in space, and even (iii) processes 
  in the Earth-Moon system or Jupiter with its satellites against the background
  of the Sun's gravity.

  Now, if we take an arbitrary slowly changing $\Phi$-dependent metric
  $q\mn$, with any positive value of $q_{00}$, it should be rescaled to
  the background value $g_{00} =1$, i.e., we must put
  $q\mn = q_{00}(\Phi_0) g\mn$, hence in a neighborhood of $x^\mu = x_0^\mu$
\beq
	Q(\Phi) = Q(\Phi_0) (1 + 2\PhiN + \ldots),             \label{Q1}
\eeq
  where we have denoted for brevity $Q(\Phi) \equiv q_{00}(\Phi)$.
  On the other hand, we have the Taylor decomposition
\beq                                                           \label{Q2}
	Q(\Phi) = Q(\Phi_0 + \Phi_1)
		= Q(\Phi_0) + Q_\Phi(\Phi_0) \Phi_1 + \ldots,
\eeq
  where the subscript $\Phi$ denotes $d/d\Phi$. Comparing (\ref{Q1}) and
  (\ref{Q2}), we obtain
\beq                                                           \label{PhiN}
	\Phi_1 =  \frac{2Q(\Phi_0)}{Q_\Phi(\Phi_0)} \PhiN.
\eeq

  The quantity $Q(\Phi) = q_{00}$ in \eq (\ref{PhiN}) and in all expressions
  where it is multiplied by the small quantity $\Phi_1$ or $\PhiN$ is
  equal to $Q(\Phi_0)$, in other words, it is a component of the slowly
  changing background metric, which in turn depends on the slowly changing
  background field $\Phi(x^\mu)$. Thus according to (\ref{q_})
\bearr                                                         \label{Q0}
	Q(\Phi) = \frac{1}{\alpha} - \frac{\beta}{\alpha + \beta}
		  \frac{(\d_t\Phi)^2}{w}
\nnn		  
	   = \frac{1}{\alpha} \biggl(1 -\frac{\beta Y}{\alpha^3 V}\biggr)
	   = \frac{1}{\alpha} \biggl[1 -\frac{4\beta Y}{(\alpha-3)^2}\biggr],
\ear
  where
\beq                                                            \label{Y}
	Y(x^\mu) \equiv \frac{(\d_t\Phi)^2}{w}
	   	 \equiv \frac{(\d_t\Phi)^2}{\eta\MN\d_\mu\Phi \d_\nu\Phi}.
\eeq

  Now we substitute the expression (\ref{Phi1}) with $\Phi_0 = \const$ into
  (\ref{eq3}) with $p=0$ assuming $\Phi_1 = \Phi_1(x^i)$, that is,
  neglecting its possible time dependence in accordance with the slow motion
  approximation. Consequently, $\Box_M\Phi = -\Delta_M \Phi_1$
  (where $\Delta_M$ is the flat-space Laplace operator corresponding to
  the metric $\eta\mn$). In the same approximation the second term with $w$
  in (\ref{eq3}) is also negligible. We thus obtain
\beq
	\Delta_M \Phi_1 =                                     \label{N1}
	\frac{\kappa \rho}{\alpha^2 (\alpha-3) V^{3/2}},
\eeq
  or, with (\ref{PhiN}),
\beq                                                          \label{N2}
	\Delta_M \PhiN = \frac{Q_\Phi}{2Q}\,
	\frac{\kappa \rho}{\alpha^2 (\alpha-3) V^{3/2}},
\eeq
  where all $\Phi$-dependent quantities on the right-hand side are taken at
  $\Phi = \Phi_0$ and $Q(\Phi)$ is specified by \eq (\ref{Q0}).

  This is still not the end of the story. The point is that the derivatives involved 
  in $\Delta_M$ are taken with respect to the  coordinates $x^i$ in which 
  the auxiliary flat metric has the form  $\eta\mn = \diag(1,-1,-1,-1)$, while 
  the physical metric $q\mn$ in the flat-space approximation (actually, the 
  tangent space metric) has the form  $q\mn = Q(\Phi_0) \eta\mn$, whereas 
  the Poisson  equation (\ref{Poisson}) expressing Newton's law should be 
  obtained in terms of the rescaled coordinates $y^i = \sqrt{Q} x^i$, i.e., the 
  spatial  coordinates corresponding to the physical metric of the form $\eta\mn$.
  Since in our approximation $Q= \const$, we can write $\d/\d x^i = \sqrt{Q}
  \d/\d y^i$, hence $\Delta_M = Q \Delta \equiv Q \Delta[y^i]$, and \eq
  (\ref{N2}) in terms of $y^i$ takes the form
\beq                                                          \label{N3}
	\Delta \PhiN = \frac{Q_\Phi}{2Q^2}\,
	\frac{\kappa \rho}{\alpha^2 (\alpha-3) V^{3/2}}.
\eeq

  Comparing (\ref{N3}) with (\ref{Poisson}), we find the following final
  expression for the effective Newtonian gravitational constant $\GN$:
\bear                                                         \label{GN}
	\GN \eql  \frac{G_0 Q_\Phi}{Q^2\alpha^2 (\alpha-3) V^{3/2}}
\nn	
	\eql  \frac{8\,G_0 Q_\Phi \alpha^{5/2}}{Q^2 (\alpha-3)^4}
		\sign(\alpha-3),
\ear
  where $V$ is taken from (\ref{func}), and $Q$ should be taken from (\ref{Q0}). 
  The factor $\sign(\alpha-3)$ in (\ref{GN}) appears because according to 
  (\ref{func})  $\sqrt{V} = |\alpha-3|/(2\alpha^{3/2})$.

  The expression (\ref{GN}) shows the following features of $\GN$ in GSG:
\begin{itemize}
\item
  	$\GN$ is variable in space and time, depending on the behavior of
	the fundamental scalar field $\Phi$;
\item
  	$\GN >0 $ only if $Q_\Phi (\alpha-3) > 0$, otherwise we obtain
	antigravity;
\item
        Due to the factor $Y$ in (\ref{Q0}), $\GN$ is different in the cases
	of cosmological and spatial dependences of $\Phi$. More
	specifically, if $\Phi = \Phi(x^0)$, we have $Y = 1$ in (\ref{Q0})
	whereas if $\Phi = \Phi(x^i)$, we obtain $Y = 0$.
\end{itemize}

\section {Variations of $\GN$}
\subsection {Cosmological variations}

  Assuming $\Phi = \Phi(x^0)$ and substituting (\ref{func}), we obtain
\bearr
	Q = \frac{4}{(\alpha - 3)^2} \ \then \                \label{Q+}
\nnn	
	\frac{Q_\Phi}{Q^2} = -\Half (\alpha-3) \alpha_\Phi
	=\alpha (\alpha-3),
\ear
  and \eq (\ref{GN}) leads to
\beq                                                         \label{GN+}
	\GN = \frac{8\,G_0 \alpha^{7/2}}{(\alpha-3)^3}\sign(\alpha-3)
	    = \frac{8\,G_0 \alpha^{7/2}}{|\alpha-3|^3}.
\eeq
  Evidently, in this cosmological setting the effective gravitational
  constant $\GN$ is always positive.

  A logarithmic derivative of $\GN$ with respect to physical time $t$
  (with $\alpha = e^{-2\Phi}$ according to (\ref{func}))
  gives its cosmological variation
\beq                                                         \label{var+}
	\frac{\dot \GN}{\GN} = \dot\Phi\,\frac{21 -\alpha}{\alpha -3},
\eeq
  where the dot denotes $d/dt$. Furthermore, in the cosmological models of
  GSG considered briefly in \cite{no12} and in detail in \cite{no14},
  $\dot\Phi$ coincides with the Hubble parameter $H$:
\beq                                                         \label{var++}
	\dot\Phi = H(t) = \frac{\dot a}{a} \ \then\
	\frac{\dot \GN}{\GN} = H\,\frac{21 -\alpha}{\alpha -3},
\eeq
  where $a(t)$ is the cosmological scale factor.

  Recalling the tight observational constraints on variations of $\GN$,
  according to which the variation (\ref{var++}) in the modern epoch should
  be smaller than the Hubble rate $H$ by at least a factor of 1000 \cite{Gt1},
  we can conclude that the cosmological models of GSG have a very small
  chance to be viable. Indeed, to fit the present-day observations, we need
\beq
	\frac{21 -\alpha}{\alpha -3} \lesssim 0.001,
\eeq
  which may be regarded as a kind of fine tuning. However, there are
  tight constraints on variations of $\GN$ in the past: for example, at the
  nucleosynthesis epoch the value of $\GN$ could not differ from the modern
  one by more than 20--30 \% \cite{Gt3}. 
  According to \cite{Gt5}, the time variation of $\GN$ between the recombination 
  time ($G_{\rm rec}$) and the present epoch ($G_0$) is constrained as 
  $G_{\rm rec}/G_0 < 1.0030\ (2\sigma)$ and 
  $G_{\rm rec}/G_0 < 1.0067\ (4\sigma)$.
 
  Meanwhile, as follows from (\ref{var++}), the field $\Phi$ evolves as $\ln (a/a_0)$, 
  $a_0 = \const$, hence $\alpha \sim a^{-2}$, and according to (\ref{GN+}), 
  $\GN$ should have changed by many orders of magnitude. This contradiction shows 
  that the GSG faces serious problems, and at least its new formulation should be
  sought for to fit the observations.

\subsection {Spatial variations}

  In the case $\Phi = \Phi(x^i)$, we have $Y =0$, and (\ref{Q0}) gives
\beq
	 Q = \frac{1}{\alpha} \ \then \                     \label{Q-}
	 \frac{Q_\Phi}{Q^2} = -\alpha_\Phi = 2\alpha.
\eeq
  whence it follows
\beq
	\GN = G_0\,\frac{16 \alpha^{7/2}}{(\alpha-3)^4}\sign(\alpha-3).
\eeq
  It follows that $\GN > 0$ only if $\alpha > 3$, otherwise there is
  antigravity instead of gravity. As to spatial variations of $\GN$, we have
\beq
	\frac{\GN'}{\GN}                                  \label{var-}
		= \Phi'\,\frac{\alpha+21}{\alpha-3},
\eeq
  where the prime denotes a derivative in any spatial direction.

  Let us look whether or not the variations (\ref{var-}) are in conflict 
  with observations. To do so, let us estimate spatial variations of $\GN$ 
  in the Solar system considering the Sun's gravity as the background slowly
  changing field. Since this field itself is rather weak, it can be taken in
  the Newtonian approximation, with the potential
\beq              \label{PhiS}
	\PhiN = \GN{}_0 M_{\odot}/r,
\eeq
  where $M_\odot$ is the solar mass, $r$ is the distance from the
  solar center, and $\GN{}_0$ is the asymptotic value of the Newtonian
  constant sufficiently far from the Sun but still not too far, so that
  the gravitational field of the Galaxy could be regarded constant at this
  scale. We denote the corresponding background value of $\Phi$ by $\Phi_0$.
  Furthermore, for $q_{00} = Q(\Phi)$ we have the expression (\ref{Q1}),
  where $Q(\Phi_0) = 1/\alpha$, and $\alpha = \e^{-2\Phi_0}$ is an unknown
  constant, for which we should only require $\alpha > 3$ to provide a
  positive value of $\GN$.

  To estimate $\GN$ variations according to (\ref{var-}), we need $\Phi' =
  d\Phi/dr = \Phi'_1$, corresponding to the expansion
  $\Phi = \Phi_0 + \Phi_1 + ...$. As follows from (\ref{PhiN}) and
  $Q = 1/\alpha$, in our case $\Phi_1 = \PhiN$, and therefore the local
  Newtonian constant changes according to
\beq                                                        \label{var-S}
	\frac{\GN'}{\GN{}_0} \approx \frac{\GN{}_0 M_{\odot}}{r^2}\,
			\frac{\alpha+21}{\alpha-3},
\eeq
  where
\beq                                                        \label{GM-S}
  	\GN{}_0 M_{\odot} \approx 1.5\ {\rm km}
\eeq
  is half the Schwarzschild gravitational radius of the Sun. The
  $\alpha$-dependent factor can be large but it tends to unity at large
  $\alpha$. In our estimates, we will put this factor equal to unity, which
  corresponds to lower limits of the corresponding variations.

  Using (\ref{var-S}), it is straightforward to find that a fractional
  variation of $\GN$ at the Earth's orbit is about $10^{-16}/{\rm km}$,
  which makes a relative difference in $\GN$ values of $\approx 10^{-12}$
  along the Earth's diameter and $\approx 0.5 \ten{-10}$ at the diameter of
  the Moon's orbit. The corresponding displacement of the lunar orbit would
  be about 4 cm, which could in principle be noticed by laser ranging.

  But even if the above variations of $\GN$ are admissible from an
  observational viewpoint, they become much larger on the planetary scale.
  Indeed, it follows from (\ref{var-S}) that the total relative variation of
  $\GN$ from Mercury's orbit to (conditional) infinity, or, say, to Kuiper's
  belt, is about $2.7\ten{-8}$ whereas an analysis of ephemerides makes it
  possible to trace annual $\GN$ variations up to $10^{-13}$ \cite{Gt1}. 
  Evidently, anomalies due to variable $G$ cannot be discovered from 
  observations of bodies in circular or near-circular orbits (for which the 
  product $\GN M$ remains constant) but should be quite detectable for 
  bodies in highly excentic orbits like some asteroids and  comets.

  For example, the asteroid Icarus has an excentric orbit located between
  approximately 2 and 0.2 astronomic units ($3\ten{12}$ to $3\ten{13}$ cm).
  Its half-revolution time is about 200 days $\sim 1.8\ten{7}$ s, while an
  anomalous acceleration due to changing $G$ would contribute about 0.1 s to
  this time (as can be roughly estimated by considering the free-fall time
  from aphelion to perihelion). At typical velocities about 30 km/s this
  corresponds
  to a displacement of 3 km, easily detectable by ranging methods but making
  problems for optical telescopes (from a distance of, say, 60 million
  kilometers, a 3-km segment is seen at an angle of 0.01'').

  We conclude that spatial $\GN$ variations in GSG, at least in its
  presently discussed formulation, do not seem to be in sharp conflict with
  observations, but are definitely in tension with them.

  It might be tempting to compare the extra accelerations of bodies due to
  changing $\GN$ with the so-called Pioneer anomaly. The latter (see the
  recent review \cite{iorio}) consists in that the radio tracking data from
  Pioneer 10 and 11 spacecrafts have revealed their constant and uniform
  deceleration approximately directed towards the Sun,
\beq                    \label{Pio}
	A_{\rm Pio} = (8.74 \pm 1.33) \times 10^{-10}\ {\rm m\,s^{-2}}
\eeq
  at heliocentric distances of 20 to 70 astronomic units (a.u.), that is,
  $(3 \div 10.5)\ten{12}$ m. The anomalous acceleration $\Delta a$ due to
  $\GN$ variation is about $10^{-8}$ times the standard Newtonian
  acceleration $a_N$ due to Sun's gravity, which is $1.5\ten{-5}\ {\rm
  m\,s^{-2}}$ at a distance of 20 a.u., hence $\Delta a \sim 10^{-13}\ {\rm
  m\,s^{-2}}$, much smaller than the Pioneer anomaly, and decreases with
  distance together with $a_N$. Thus a variable $G$ cannot account for the
  Pioneer anomaly, which is, according to \cite{iorio}, consistent with
  known physics, being almost completely explained by thermal radiation from
  the spacecrafts.

\section{Conclusion}

  We have to conclude that the cosmological variations of the Newtonian
  gravitational constant predicted by the GSG are in striking conflict
  with observations whereas its spatial variations are comparatively small
  (at least in the Solar system) but are easily detectable by modern
  instruments.

  A point of interest is an apparent contradiction between the geodesic
  nature of test body paths in the metric $q\mn$ in GSG and the existence of
  anomalous accelerations due to varying $\GN$ in the Newtonian
  approximation. A possible explanation is that the Newtonian approximation
  deals with small velocities and gravitational potentials, and the
  anomalous acceleration exists in the next order of magnitude with respect
  to the flat-space approximation. In other words, the true geodesics are slightly 
  non-Newtonian and are better described in the post-Newtonian approximation.  
  The same is true for scalar-tensor and $f(R)$ theories of gravity.
  
  The present results have been obtained for the particular version of GSG 
  presented in \cite{no12, no14, no16, toni19}. It would be of interest to 
  find out how they can change in its more general formulations discussed in 
  \cite{shi15, shi16, jardim15}, which can be a subject for future studies.

\subsection*{Acknowledgments}

  I am grateful to Milena Skvortsova, Sergei Bolokhov, Mario Novello,
  Julio C. Fabris and Junior D. Toniato for helpful discussions.
  This publication was supported by the RUDN University program 5-100.
  The work was also performed within the framework of the Center FRPP 
  supported by MEPhI Academic Excellence Project 
  (contract No. 02.a03.21.0005, 27.08.2013).

\end{document}